\documentclass{article}

\usepackage{arxiv}

\usepackage[utf8]{inputenc} 
\usepackage[T1]{fontenc}    
\usepackage{hyperref}       
\usepackage{url}            
\usepackage{booktabs}       
\usepackage{amsfonts}       
\usepackage{nicefrac}       
\usepackage{microtype}      
\usepackage{lipsum}
\usepackage{graphicx}
\usepackage{graphicx}
\usepackage[table,xcdraw]{xcolor}
\usepackage{makecell}
\usepackage{threeparttable}
\usepackage{float}
\usepackage{multirow}
\usepackage[numbers]{natbib}
\usepackage{amsmath}
\graphicspath{ {./images/} }

\title{Substrate Effect on Electronic Band Structure and Topological Property in Monolayer V$_2$O$_3$ Magnetic Topological Insulator}

\author{
 Zheng Wang$^{1,\#}$, Jingshen Yan$^{1,\#}$, Shu-Shen Lyu$^{1-3}$, Kaixuan Chen$^{1-3,*}$ \\
  $^1$School of Materials, Shenzhen Campus of Sun Yat-sen University, Shenzhen, 518107, PR China,\\
  $^2$Guangdong Engineering Technology Research Centre \\for Advanced Thermal Control Material and System Integration (ATCMSI),\\ Sun Yat-sen University, Guangzhou, 510275, PR China,\\
  $^3$Huizhou Research Institute, Sun Yat-sen University, Huizhou, 516081, PR China.\\
  \#These authors contribute equally to this work.\\
  \texttt{E-mail: chenkx26@mail.sysu.edu.cn} \\
   \\
}

\begin{document}
\maketitle
\begin{abstract}
Monolayer V$_2$O$_3$, a two-dimensional magnetic topological insulator with intrinsic ferromagnetic order and a nontrivial band gap, offers a promising platform for realizing quantum anomalous Hall (QAH) states. Using first-principles density functional theory calculations, we systematically investigate the impact of substrate selection on its electronic and topological properties. By modeling heterostructures with van der Waals (vdW) substrates, we demonstrate that non-magnetic substrates such as h-BN preserve the QAH phase with a Chern number C = 1, maintaining gapless chiral edge states. In contrast, ferromagnetic substrates induce extra electrons, destroying the topological order by shifting the Fermi level. These findings establish substrate engineering as a pivotal strategy for experimental realization of dissipationless edge transport in V$_2$O$_3$-based vdW heterostructures, advancing their potential applications as low-power topological electronics.
\end{abstract}


\section{Introduction}
Topological insulators (TIs), a novel class of quantum materials, have garnered significant attention because of their unique properties, which exhibit bulk insulating while simultaneously possessing conductive states on their surface (or edges)\cite{WANG2021100098,changColloquiumQuantumAnomalous2023,denner2021exceptional}. Among them, as recently predicted theoretically, two-dimensional (2D) honeycomb lattice V$_2$O$_3$ was proposed as a candidate for a 2D magnetic topological insulator, emerging potential to host quantum anomalous Hall (QAH) states due to its experimental feasibility, intrinsic ferromagnetism, and robust band gap\cite{surnevGrowthStructureUltrathin2000,wangPredictionHightemperatureQuantum2017}. 

However, the 2D materials must be supported on substrates, whose interaction with the film can significantly modify its electronic and topological features\cite{wang2025effects,yan2024exchange,yangElectronicStructuresMagnetic2022}. In particular, substrate-induced symmetry breaking or magnetic proximity effects may jeopardize the QAH phase\cite{lin2013substrate,cardoso2023strong}. 

For $d$-orbital-mediated magnetic topological insulators, the circumstantial issue of substrate selection, which influences or disrupts the topological character of V$_2$O$_3$, still remains elusive. In this work, we systematically investigate the impact of both magnetic and non-magnetic substrates on the electronic structure and topological properties of V$_2$O$_3$ film using first-principles calculations. We show that while non-magnetic insulator substrates preserve the QAH state, ferromagnetic substrates disrupt the topological order due to the extra magnetic moments induced metallic state of system. Our findings demonstrate the crucial role of substrate selection in realizing and controlling topological phases in 2D magnetic TIs.

\section{Results and discussion}

\begin{figure}[ht]
\centering
  \includegraphics[width=0.5\linewidth,scale=1.00]{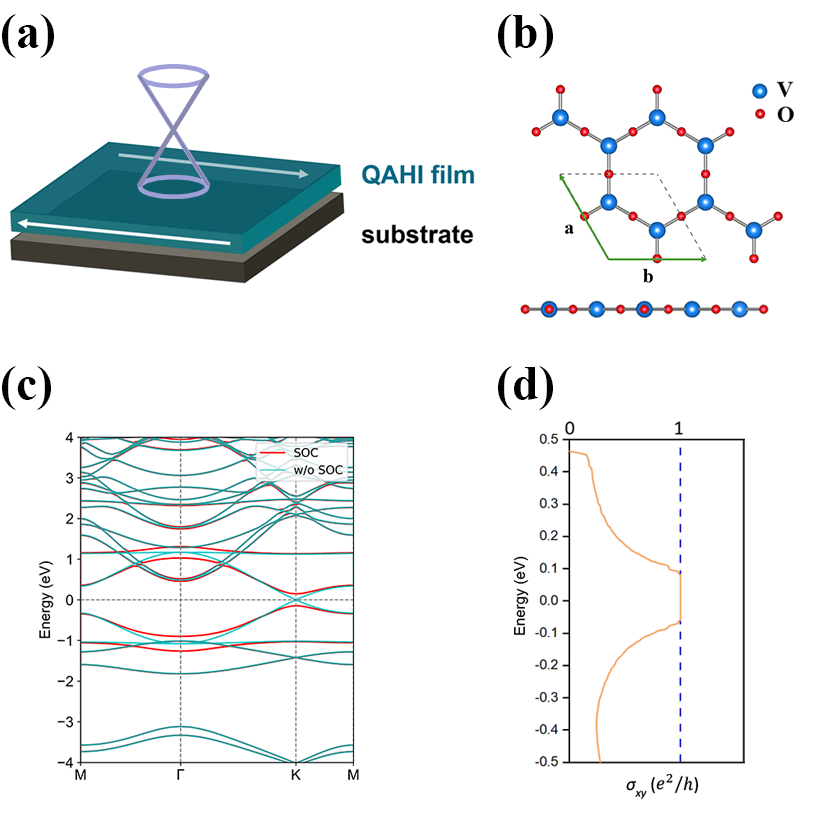}
  \caption{(a) Schematic demonstration of heterostructure, i.e., V$_2$O$_3$ supported on substrates. (b) Atomistic structure of monolayer V$_2$O$_3$ (top view and side view). (c) Electronic band structure of monolayer V$_2$O$_3$ with and without SOC effect. (d) The corresponding quantized QAH conductance.}
  \label{fig0}
\end{figure}

To investigate the influence of substrates on the topological properties of V$_2$O$_3$ thin films (the electronic properties are shown in Fig.\ref{fig0}), we selected several 2D materials exhibiting good lattice matching with V$_2$O$_3$ to construct simplified heterostructures. The spin-orbit coupling (SOC)-induced band splitting of the V$_2$O$_3$ film was labeled as the topological index.

\begin{table}[h!]

\caption{Structural information of the constructed V$_{2}$O$_{3}$-substrates heterostructures.}
\label{tableb1}
\resizebox{\columnwidth}{!}{%
\begin{tabular}{cccc|ccccc}
\hline
Substrates & \makecell[c]{Space \\ Group} &\makecell[c]{Lattice \\ constant (Å)} & \makecell[c]{Band\\ Gap (eV)}& \makecell[c]{Stacking configuration \\ (V$_{2}$O$_{3}$/substrates)} & \makecell[c]{Interlayer \\ spacing (Å)} & \makecell[c]{Lattice \\ Mismatch ($\%$)} & \makecell[c]{Heterostructures \\ parameters (Å)}\\  \hline
CoCl$_{2}$\cite{botana2019electronic}&P$\Bar{3}$m1& 3.54 &0.08&1×1/$\sqrt{3}\times\sqrt{3}$ & 2.8 & 0.5 &a=b=6.18\\
FeCl$_{2}$\cite{10.1063/1.4921096}&P$\Bar{3}$m1& 3.57 &0 &1×1/$\sqrt{3}\times\sqrt{3}$ & 3.2 & 0.1 &a=b=6.18\\
NiCl$_{2}$\cite{kulishSinglelayerMetalHalides2017}&P$\Bar{3}$m1& 3.48 &0&1×1/$\sqrt{3}\times\sqrt{3}$ & 3.4 & 2.6 &a=b=6.18\\
MgCl$_{2}$\cite{mahida2022first}&P$\Bar{3}$m1& 3.64 &5.90&1×1/$\sqrt{3}\times\sqrt{3}$ & 3.2 & 2.0 &a=b=6.19\\
CrCl$_{3}$\cite{cai2019atomically} &R$\Bar{3}$& 6.00 &1.52 &1×1/1×1 & 2.9 & 2.9 &a=b=6.14\\
WTe$_{2}$\cite{li2020synthesis}  & P6$_3$mmc & 3.56 &0.94 &1×1/$\sqrt{3}\times\sqrt{3}$ & 3.3 & 0.2 &a=b=6.13\\
h-BN\cite{satawara2021structural}  &R$\Bar{3}$& 2.50 &4.90 &2×2/5×5 & 3.5 & 1.1 &a=b=12.51\\ \hline
\end{tabular}%
}
\end{table}

\begin{figure}[h!]
    \centering
    \includegraphics[width=\linewidth,scale=1]{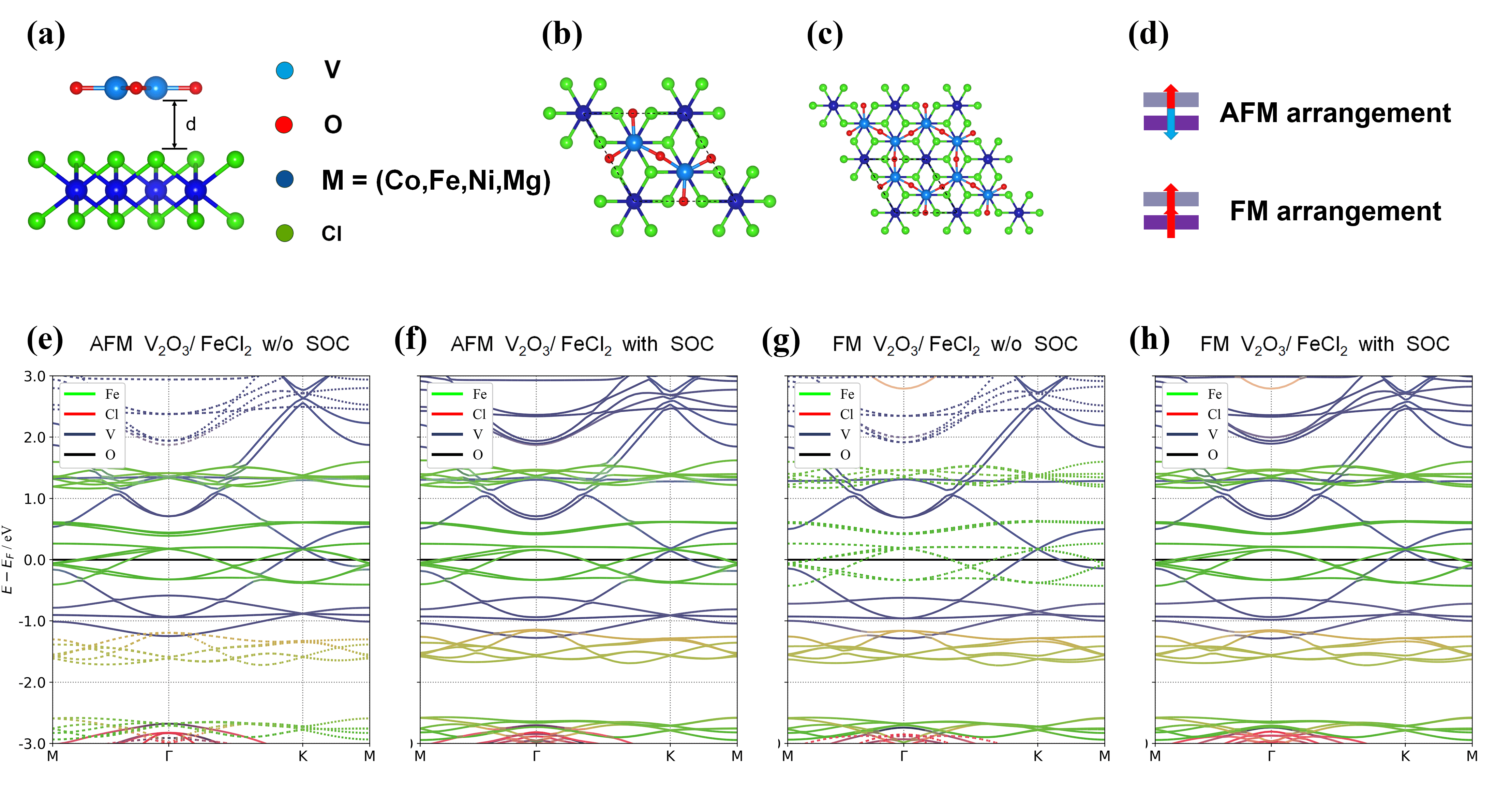}
    \caption{(a) top view, (b) side view, (c) supercell perspective of V${_2}$O${_3}$/MCl${_2}$ heterostructures. (d) magnetic configurations of V${_2}$O${_3}$/MCl${_2}$ heterostructures. (e) and (f) band structures of AFM V${_2}$O${_3}$/FeCl${_2}$ with/without SOC. (g) and (h) band structures of FM V${_2}$O${_3}$/FeCl${_2}$ with/without SOC.}
    \label{fig1}
\end{figure}

The corresponding substrates are listed in Table\ref{tableb1}, wherein MCl$_2$ (M=Fe, Co, Ni), CrCl$_3$ are $d$-orbital-mediated magnetic materials, while MgCl$_2$, h-BN, and WTe$_2$ are non-magnetic insulator materials. The heterostructures were constructed by directly stacking the V$_2$O$_3$ film onto the selected substrates along the $z$-axis, ensuring a lattice mismatch of less than 5$\%$ for all systems. 

Given the prevalent use of magnetic substrates to drive topological insulator phase transitions\cite{liuMagneticTopologicalInsulator2023}, we initially investigated whether the magnetic proximity effect induced by van der Waals layers would influence the topological properties of V$_{2}$O$_{3}$. Co(Fe,Ni)Cl$_{2}$ monolayers are employed as substrates since they have high lattice matching and were synthesised experimentally before\cite{botana2019electronic,10.1063/1.4921096,kulishSinglelayerMetalHalides2017}. The optimized heterostructures are shown in Fig.\ref{fig1}(a, b, c) presenting the side/top view and supercell, respectively. The V atoms in the upper layer align directly with the lower M (Fe,Co,Ni,Mg) atoms along the $z$-axis. The crystal structure exhibits a van der Waals layered framework, with no obvious lattice distortion observed from the supercell perspective. Furthermore, we considered both AFM and FM coupling between the film and substrates to investigate its effect, as shown in Fig.\ref{fig1}(d). 

\begin{figure}[h!]
    \centering
    \includegraphics[width=\linewidth,scale=1.00]{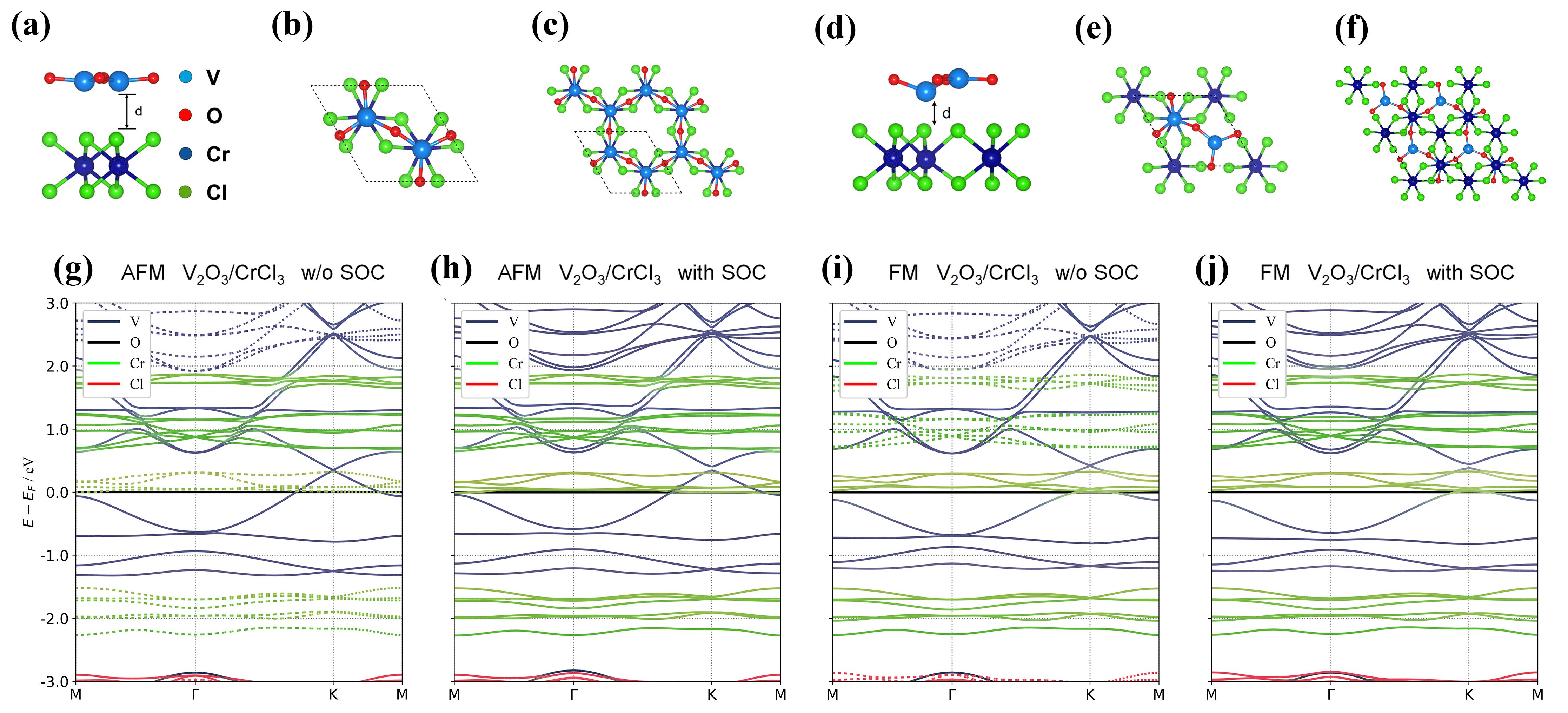}
    \caption{(a-c)/(d-f) side view, top view and supercell perspective of AA/AB stacking V${_2}$O${_3}$/CrCl${_3}$ heterostructures. (g) and (h) band structures of AFM AA stacking V${_2}$O${_3}$/CrCl${_3}$ with/without SOC. (g) and (h) band structures of FM AA stacking V${_2}$O${_3}$/CrCl${_3}$ with/without SOC.}
    \label{fig2}
\end{figure}

As reported in previous studies, the topological properties of V${_2}$O${_3}$ originate from a Dirac point (contributed by V $d_{xz}$,$d_{yz}$ orbitals) at the high symmetry $K (K')$-point, which arises due to a $C_{3v}$ symmetry-protected band crossing, in combination with band inversion induced by the strong SOC effect of the $d$-orbital electrons\cite{wangPredictionHightemperatureQuantum2017}. Given the strong correlation between the electronic band structure and topological properties, we performed spin-polarized and SOC calculations on V${_2}$O${_3}$/MCl${_2}$ heterostructures. It should be noted that the MX$_{2}$ primitive cell has a lattice constant of approximately 3.5Å, when combining a 1×1 V$_{2}$O$_{3}$ unit cell with a 2×2 MX$_{2}$ cells resulting in a lattice mismatch exceeding 10$\%$, which would introduce significant lattice distortion. Therefore, $\sqrt{3}\times\sqrt{3}$ MX$_{2}$ substrates were employed (see Fig.\ref{fig2}a-c). In the AFM coupling system, as illustrated in Fig.\ref{fig1}(e), the spin-up bands near the Fermi level orbitals are primarily contributed by V (blue line) and Fe (green line). At the $K$-point, the Dirac cone of the V bands overlaps with the Fe $d$ orbitals. This hybridization occurs between orbitals of the same spin and energy, which can also be observed from the perspective of the wavefunction at the $K$ point correspondingly in Fig.S1(a,b). The spin-up wave function of the Dirac cone slightly distributes on the lower FeCl${_2}$ layer, dominated by the $d-d$ hybridization between V and Fe. When SOC is considered, the band structure (Fig.\ref{fig1}(f)) and the corresponding wavefunctions exhibit negligible changes. The Dirac cone remains almost gapless under SOC, thereby destroying the topological nature of the V$_2$O$_3$ film. In the FM counterpart shown in Fig.\ref{fig1}(g), the positions of the spin-up and spin-down FeCl${_2}$ bands are reversed, thus the bands hybridization at $K$-point in AFM are decoupled by flipped magnetic moments. Moreover, the wavefunctions at the band crossing are not contributed by the Fe $d$ orbitals. When SOC is included (Fig.\ref{fig1}(h)), a tiny SOC splitting appears at the Dirac cone of V${_2}$O${_3}$, as same as the AFM configuration. The SOC splitting energies are listed in Table\ref{tableb2}, we find the magnetic coupling configuration appears to have no significant influence on the SOC induced splitting energy.

\begin{table}[ht]
 \centering
  \caption{Calculated SOC splitting energy (in meV) of V$_2 $O$_3 $ supported on different substrates under AFM/FM magnetic configuration.}
  \label{tableb2}
  \resizebox{0.45\columnwidth}{!}{%
  \begin{tabular}{ccccccc}
    \hline
Substrates&\multicolumn{2}{c}{CoCl$_{2}$} & \multicolumn{2}{c}{FeCl$_{2}$}&\multicolumn{2}{c}{NiCl$_{2}$}\\ \hline
\makecell[c]{Magnetic\\Configuration} &AFM&FM&AFM&FM&AFM&FM \\ \hline
\makecell[c]{Splitting\\ energy} & 35        & 38       & 40        & 38       & 42        & 37      \\ \hline
  \end{tabular}%
  }
\end{table}

Besides, magnetic substrates introduce magnetic moments arising from extra unpaired electrons, effectively resulting in an electron-doped system. This doping shifts the Fermi level, thereby destroying the topological nature of the system. But it is still necessary to investigate the mechanism of broken topology, whether correlated with other magnetic substrates or not. We further employ 2D magnetic insulator CrCl${_3}$, which possesses $R\overset{-}3$ point group allowing different heterostructure stacking configurations with V${_2}$O${_3}$ film, AA and AB\cite{morosin1964x}. Fig.\ref{fig2}(a-f) present the top view, side view, supercell view of AA and AB stacking system. The AA stacking heterostructure is shown in Fig.\ref{fig2}(a-c), the upper V atoms are directly align with the lower Cr atoms, remaining 2.9Å layer distance; the hexagonal structure of V${_2}$O${_3}$ highly overlaps with the hexagonal structure of lower CrCl${_3}$ layer (Fig.\ref{fig2}(c)). While for the AB stacking system, half of the V atoms are positioned above the vacancy site in the lower CrCl${_3}$ layer. The optimized lattice exhibits severe lattice distortion along $z$-axis, so we excluded it from further analysis. Analogous to the band structures of V${_2}$O${_3}$/MCl${_2}$ discussed above, the Fermi level near the two magnetic configurations(AFM and FM) is mainly occupied by V and Cr, as shown in Fig.\ref{fig2}(g,i). In FM system, the $d-d$ orbital overlapping induces significant hybridization, rendering discontinuous splitting of the V bands near the band crossing. When SOC is considered, as presented in Fig.\ref{fig2}(h,j), the SOC induced splitting energy of 63 meV occurs in both configurations at the band crossing. The corresponding wavefunctions in real space are posted in Fig.S1(c,d). It should be noted that the magnetic coupling configuration does not significantly influence the SOC-induced band splitting in V${_2}$O${_3}$. The observed differences in SOC splitting energy between V${_2}$O${_3}$/MCl${_2}$ and V${_2}$O${_3}$/CrCl${_3}$ appear to be primarily governed by the degree of band overlap or the density of states at the overlapping regions, which are shown in Fig.S4. 

Subsequently, we investigate the effect of interlayer spacing on the SOC-induced band splitting in V${_2}$O${_3}$ thin films by explicitly varying the interlayer distance within the heterostructures. Notably, an excessively small interlayer spacing can induce strong bonding between V${_2}$O${_3}$ and CrCl${_3}$ under AFM coupling; we then focus solely on the FM coupling. As listed in Table\ref{tableb3}, the splitting energy increases with increasing interlayer spacing, exhibiting a clear correlation between these two parameters. When the interlayer spacing is set to 2.5 Å, the band structure of V${_2}$O${_3}$ film is significantly altered by strong orbital hybridization, the band crossing disappears. The corresponding electronic structures and wave functions are shown in Fig.S3. 

\begin{table}[ht]
 \centering
  \caption{Calculated SOC splitting energy (in meV) of V$_2 $O$_3$ supported on CrCl$_{3}$ substrate under FM coupling with different interlayer distance.}
  \label{tableb3}
  \resizebox{0.5\columnwidth}{!}{%
  \begin{tabular}{cccccc}
    \hline
Interlayer spacing (Å)         & 2.5 & 3.0  & 3.5 & 4.0  & 4.5 \\ \hline
Splitting energy (meV) & N/A & 66 & 76  & 80 & 83  \\ \hline
  \end{tabular}%
  }
\end{table}

\begin{figure}[h!]
    \centering
    \includegraphics[width=\linewidth,scale=1.00]{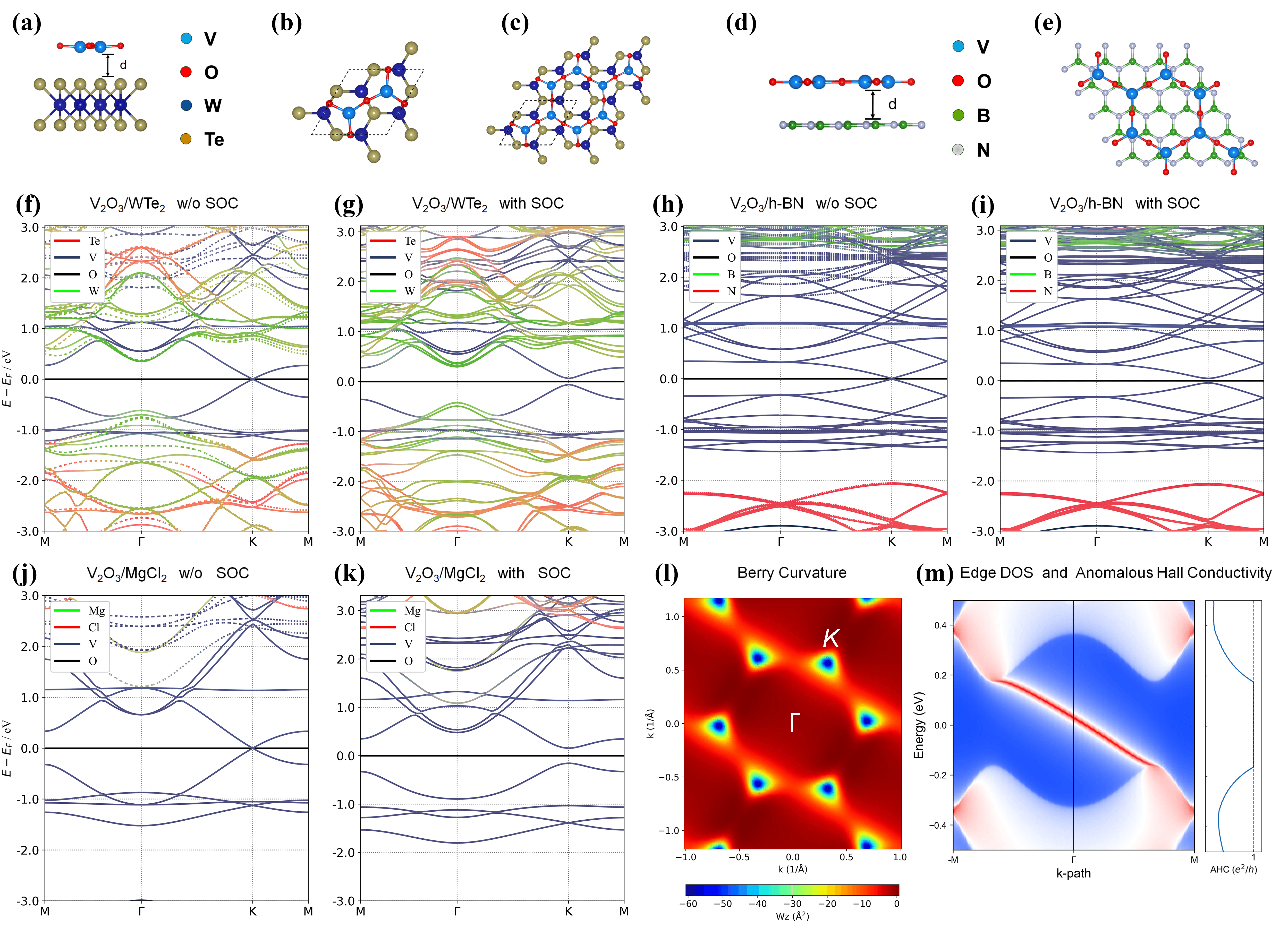}
    \caption{(a-c) side view, top view and supercell perspective of V$_2$O$_3$/WTe$_2$ heterostructure. (d,e) side view and top view of V$_2$O$_3$/h-BN heterostructure. (f) and (g) band structures of V$_2$O$_3$/WTe$_2$ with/without SOC. (h) and (i) band structures of V$_2$O$_3$/h-BN with/without SOC. (j) and (k) band structures of V$_2$O$_3$/MgCl$_2$ with/without SOC. (l) and (m) non-zero berry curvature, chiral edge state and corresponding anomalous Hall conductivity of V$_2$O$_3$/MgCl$_2$.}
    \label{fig3}
\end{figure}

Regarding non-magnetic substrates, we selected monolayer h-BN, a widely used experimental substrate; along with non-magnetic MgCl${_2}$ and WTe${_2}$, a material with strong SOC driven by heavy elements and broken inversion symmetry\cite{satawara2021structural,yanaki1973preparation,mahida2022first}. The optimized lattice structure of the V$_2$O$_3$/WTe$_2$ heterostructure is shown in Fig.~\ref{fig3}(a,b). Differently from the previously discussed MCl${_2}$ systems, each V atom in this system is aligned directly above a hollow site of the lower WTe$_2$ hexagon. Fig.~\ref{fig3}(c) presents a supercell view of the heterostructure, where the reshaped lattice constant slightly decreases to 6.13 Å, introducing a small twist in the V$_2$O$_3$ film. The corresponding band structures are shown in Fig.\ref{fig3}(f,g), the strong $d-d$ orbital hybridization occurs away from the Fermi level; due to the non-magnetic substrate, the parts of V$_2$O$_3$ bands basically retain the characteristics as a single layer, maintaining the linear Dirac cone at the $K$-point. Besides, the WTe$_2$ is introduced a spin polarization splitting, reaching up to 70 meV. When considering SOC effect, as shown in Fig.\ref{fig3}(g), the Dirac cone originating from the V$_2$O$_3$ layer opens a band gap of approximately 120 meV. Additionally, the bands contributed by the WTe$_2$ substrate exhibit significant SOC splitting near the high-symmetry $\Gamma$ and $M$ points, reflecting the intrinsic strong SOC of WTe$_2$. However, this substrate-induced SOC effect does not constructively interact with the on-site SOC of the V$_2$O$_3$. 

Fig.~\ref{fig3}(d,e) show the side and top views of the V$_2$O$_3$/h-BN heterostructure, The V$_2$O$_3$ structure exhibits no significant change when deposited on the h-BN substrate. The corresponding band structure are shown in Fig.~\ref{fig3}(h,i), the band crossing of V$_2$O$_3$ is also located at the Fermi level, and is introduced 89meV band gap by SOC effect. The optimized structure of the V$_2$O$_3$/MgCl$_2$ heterostructure is shown in Fig.\ref{fig1}(a), and the corresponding band structures are presented in Fig.\ref{fig3}(j,k). Due to the electronic configuration of Mg (outermost 3$s^2$ electrons), there is only minor orbital overlap between the Mg $s$ orbitals and Cl $p$ orbitals near the V$_2$O$_3$ bands. As shown in Fig.\ref{fig3}(k), upon including SOC, the V$_2$O$_3$/MgCl$_2$ heterostructure exhibits a remarkable SOC-induced band gap of approximately 310 meV, which is larger than that observed on other substrates.

To determine whether the SOC gap retains a topological nontrivial feature, we constructed a tight-binding (TB) model of the heterostructure based on Wannier90 and calculated the Berry curvature, edge states, and anomalous Hall conductivity. The results confirm that the opened SOC gap remains topologically nontrivial. As shown in Fig.\ref{fig3}(l), a nonzero Berry curvature is concentrated near the $K$ point in the Brillouin zone, contributing to the nontrivial topology.  Fig.\ref{fig3}(m) displays the chiral edge states crossing the Fermi level and the inset on the right side shows the anomalous Hall conductivity, where a quantized conductivity plateau near the Fermi level indicates a Chern number of $C = 1$.

\section{Conclusion}

In summary, we have systematically studied the influence of substrate selection on the electronic structure and topological properties of monolayer V$_{2}$O$_{3}$ using first-principles calculations. Our results reveal that the topological band gap of V$_{2}$O$_{3}$ is directly correlated with the magnitude of orbital overlapping and interlayer spacing; non-magnetic substrates preserve the QAH phase with a Chern number of $C=1$, while ferromagnetic substrates disrupt the TI feature via extra magnetic moments shifting Fermi level. These findings underscore the critical role of substrate engineering in stabilizing and tuning topological phases in two-dimensional magnetic materials.

\bibliographystyle{plainnat}
\bibliography{rsc} 
\nocite{*}

\end{document}


\maketitle

\section{Computational details}
\noindent First-principles calculations were performed using density functional theory (DFT) as implemented in the Vienna Ab initio Simulation Package (VASP)\cite{kresseEfficientIterativeSchemes1996}. The interactions between electrons and ions were treated using the projector augmented wave (PAW) method\cite{blochlProjectorAugmentedwaveMethod1994}. The exchange-correlation potential was described using the generalized gradient approximation (GGA) in the Perdew-Burke-Ernzerhof (PBE) form\cite{perdewGeneralizedGradientApproximation1996}. A plane-wave basis set with a kinetic energy cutoff of 500 eV was employed, and the Brillouin zone was sampled using an 11 × 11 × 1 Gamma-centered $k$-mesh. Self-consistent calculations were considered converged when the total energy difference was less than 10$^{-6}$ eV, while atomic structures were optimized until the residual Hellmann-Feynman forces were below 0.01 eV/Å. To avoid spurious interactions due to periodic boundary conditions, a vacuum layer of 15 Å was added along the $z$-axis. The DFT+$U$ approach was used to account for the on-site Coulomb interactions in the localized $d$ orbitals, with Hubbard $U$ parameters of 3.2 eV for vanadium only\cite{}. Furthermore, vdW interactions were corrected using the DFT-D3 method proposed by Grimme\cite{10.1063/1.3382344}.

The topological characteristic is confirmed by SOC-induced band splitting, berry curvature, edge state and AHC. All topology related calculations are based on Maximally Localized Wannier Functions (MLWFs) methods, as implemented in the Wannier90 and WannierTools packages\cite{mostofiUpdatedVersionWannier902014, wuWannierToolsOpensourceSoftware2018}. The Chern number is calculated by integrating Berry curvature over the first Brillouin Zone:

\begin{equation}
C=\frac{1}{2\pi}\sum_{n}\int_{\mathrm{BZ}}\ d^2k\Omega_n
\end{equation}

Where $\Omega(k)$ is the Berry curvature which is calculated by Kubo formula\cite{thouless1982quantized,yao2004first}:
\begin{equation}
\Omega_n\left(k\right)=-\sum_{n^\prime\neq n}\frac{2\mathrm{Im}\left\langle\psi_{nk}\left|\upsilon_x\right|\psi_{n^\prime k}\right\rangle\left\langle\psi_{n^\prime k}\left|\upsilon_y\right|\psi_{nk}\right\rangle}{\left(\varepsilon_{n^\prime}-\epsilon_n\right)^2}
\end{equation}

\newpage
\section{Complement to figures}

\textbf{Wavefuctions at $K$ point in real space of V$_{2}$O$_{3}$/MCl$_{2}$ heterostructures (M=Co,Ni).} 
\begin{figure}[htbp]
    \centering
    \includegraphics[width=\linewidth,scale=1]{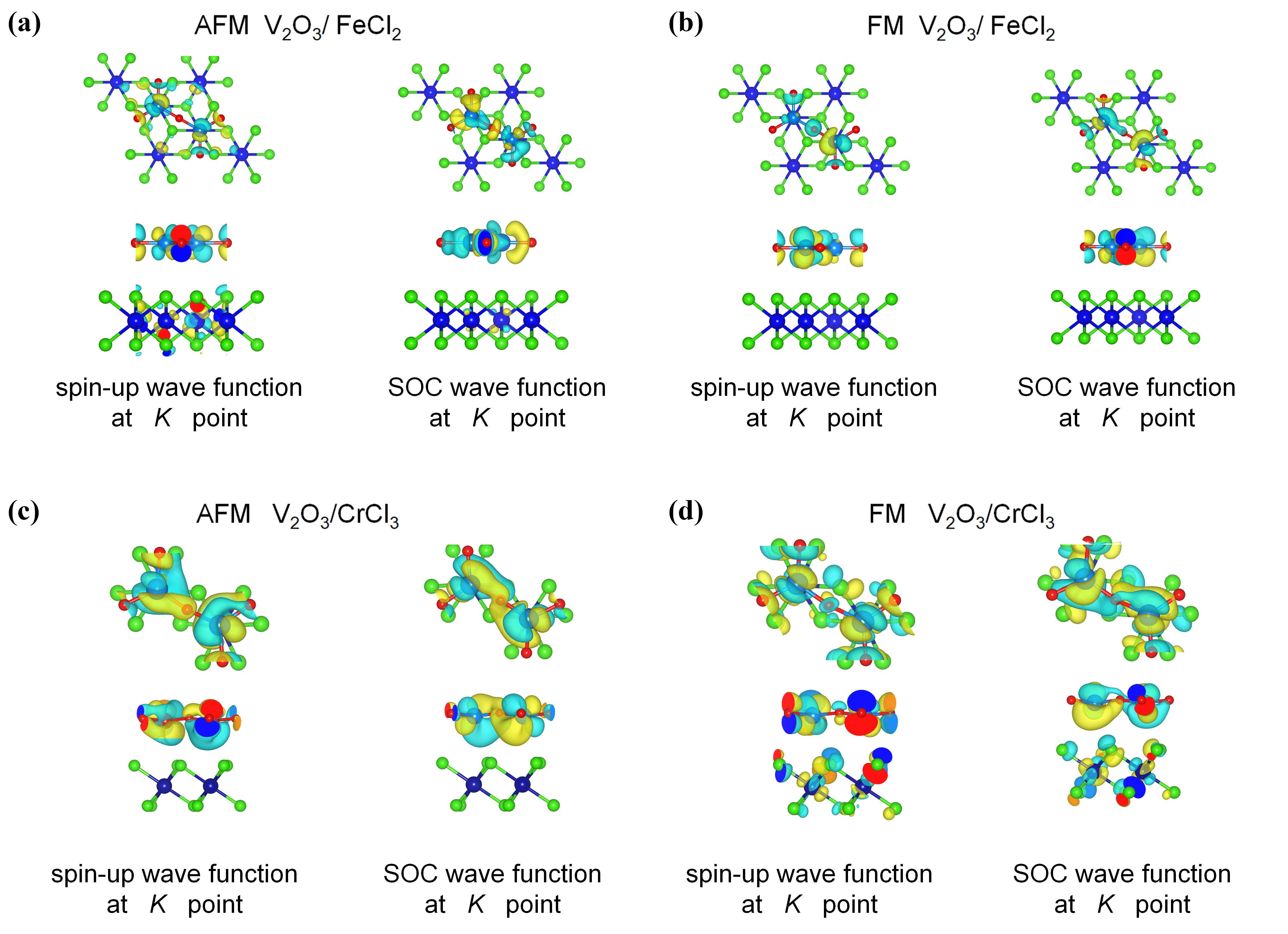}
    \caption{(a)/(b) spin-up/SOC wavefunction of V$_{2}$O$_{3}$/MCl$_{2}$ heterostructure under AFM coupling. (c)/(d) spin-up/SOC wavefunction of V$_{2}$O$_{3}$/MCl$_{2}$ heterostructure under FM coupling.}
    \label{figs1}
\end{figure}

\clearpage
\textbf{Band structure of V$_{2}$O$_{3}$/MCl$_{2}$ heterostructures (M=Co,Ni).} 
\begin{figure}[h]
    \centering
    \includegraphics[width=\linewidth,scale=1]{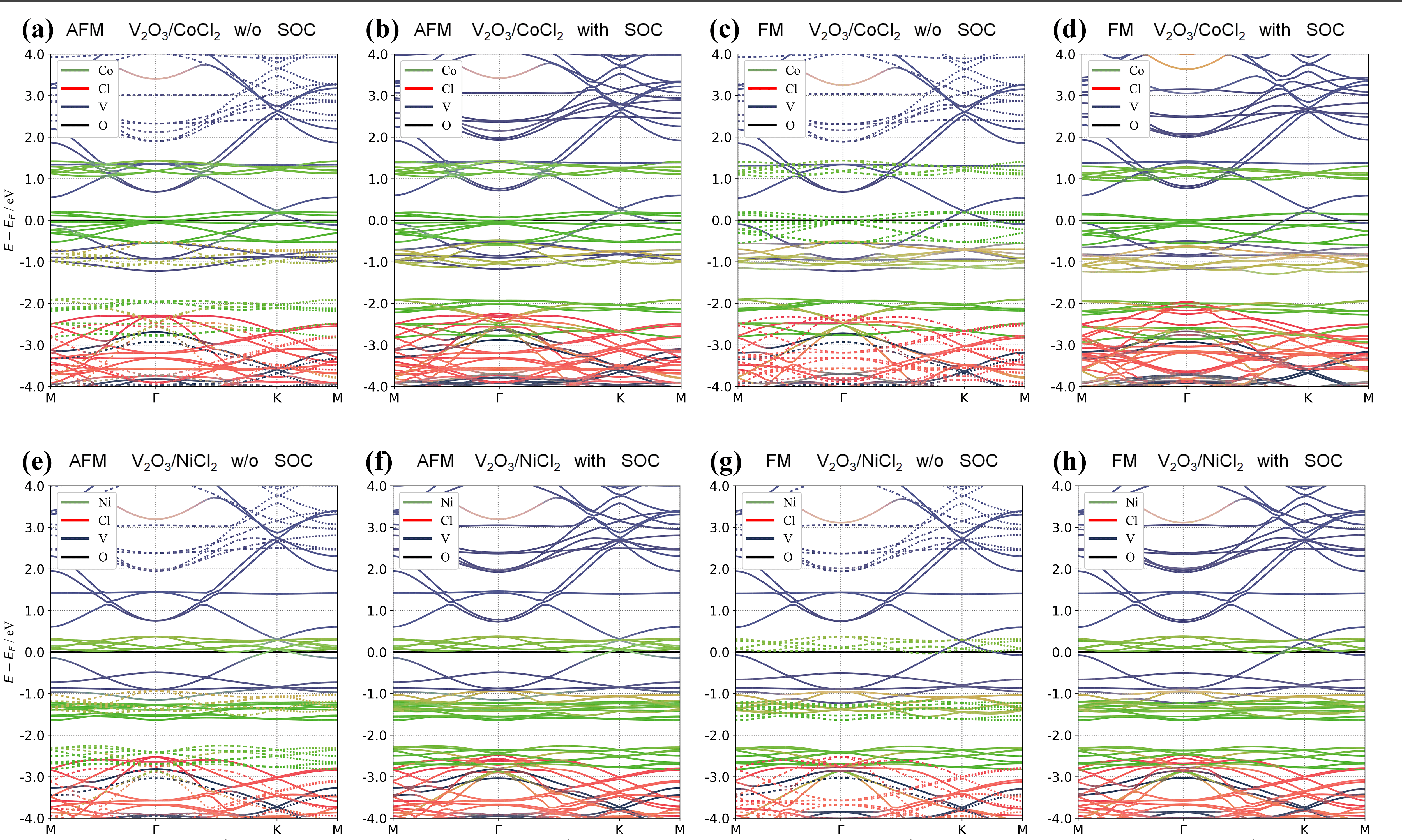}
    \caption{(a) and (b) band structures of AFM V${_2}$O${_3}$/CoCl${_2}$ with/without SOC. (c) and (d) band structures of FM V${_2}$O${_3}$/COCl${_2}$ with/without SOC. (e) and (f) band structures of AFM V${_2}$O${_3}$/NiCl${_2}$ with/without SOC. (g) and (h) band structures of FM V${_2}$O${_3}$/NiCl${_2}$ with/without SOC.}
    \label{figs2}
\end{figure}

\clearpage

\textbf{Band structures of AA stacking V$_{2}$O$_{3}$/CrCl$_{3}$ heterostructures with different interlayer spacings under AFM coupling.} 
\begin{figure}[h!]
    \centering
    \includegraphics[width=\linewidth,scale=1]{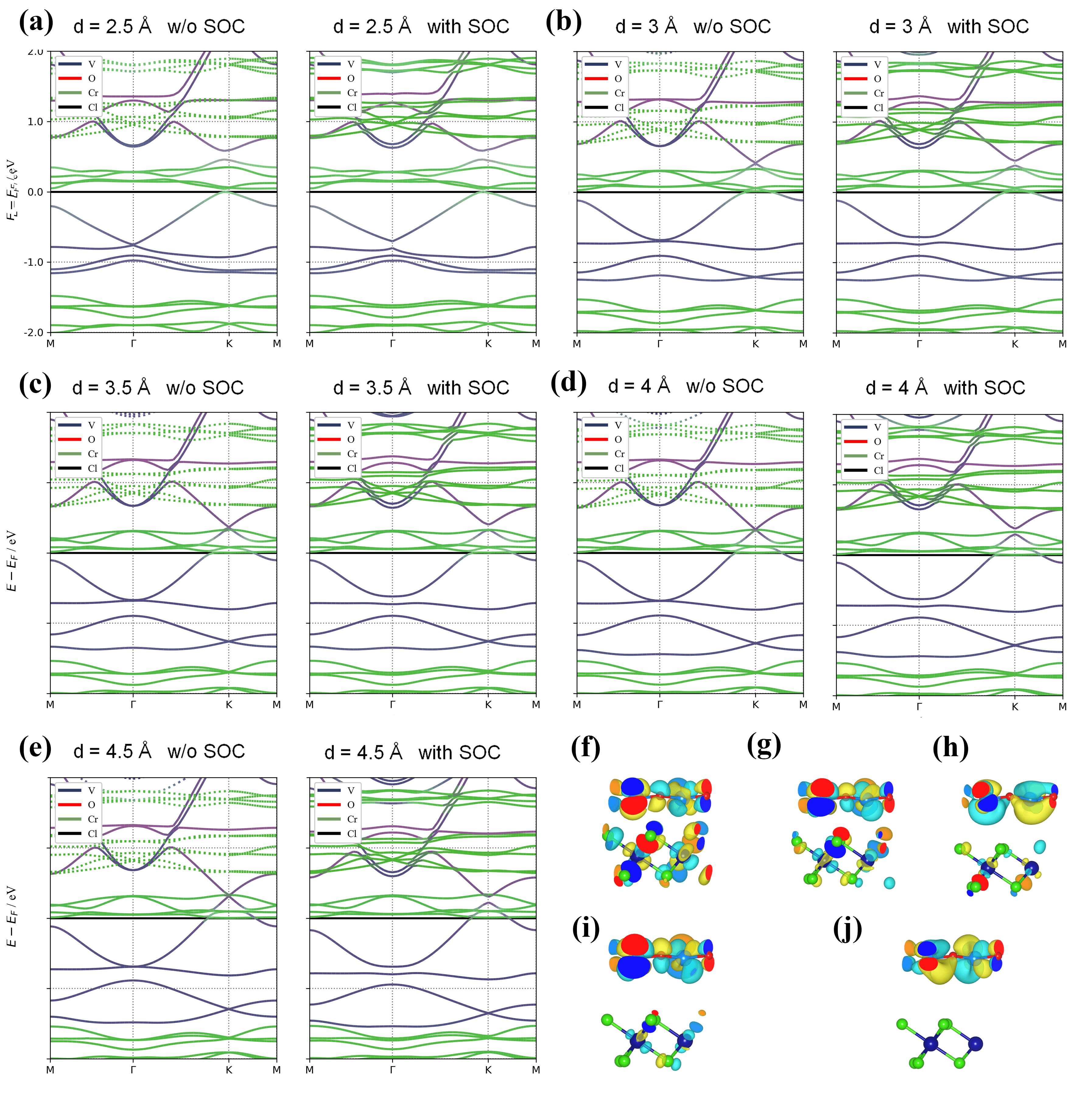}
    \caption{(a-e)spin polarization and SOC band diagram with interlayer spacing of 2.5 Å -4.5 Å. (f-j) real space wavefunctions distribution of band crossing points at $K$ point corresponding to interlayer spacing of 2.5 Å -4.5 Å.}
    \label{figs3}
\end{figure}

\clearpage
\textbf{Density of states for all heterostructures.} 
\begin{figure}[h]
    \centering
    \includegraphics[width=\linewidth,scale=1]{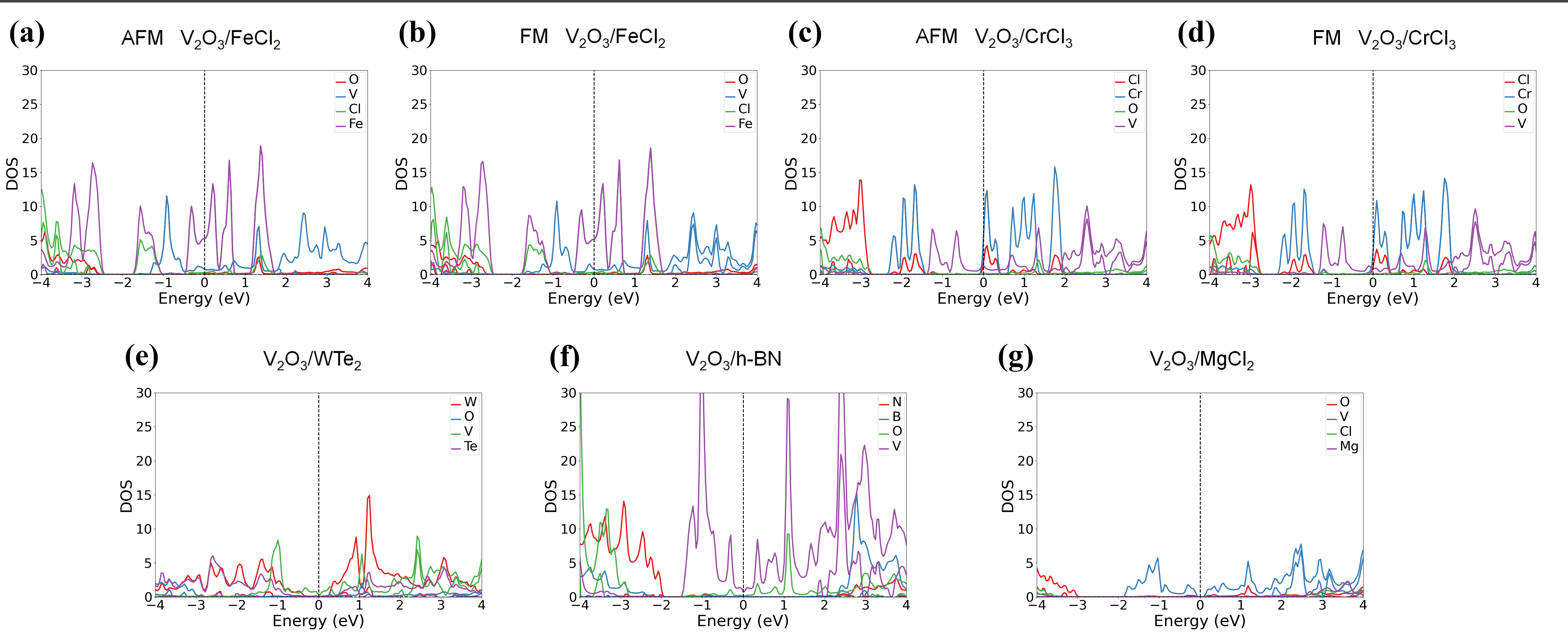}
    \caption{(a) V$_{2}$O$_{3}$/FeCl$_{2}$ under AFM coupling. (b) V$_{2}$O$_{3}$/FeCl$_{2}$ under FM coupling. (c) V$_{2}$O$_{3}$/CrCl$_{3}$ under AFM coupling. (d) V$_{2}$O$_{3}$/CrCl$_{3}$ under FM coupling. (e) V$_{2}$O$_{3}$/WTe$_{2}$. (f) V$_{2}$O$_{3}$/h-BN. (g) V$_{2}$O$_{3}$/FeCl$_{2}$}
    \label{figs4}
\end{figure}

\bibliographystyle{plainnat}
\bibliography{rsc} 